# Globalization of Scientific Communication: Evidence from authors in academic journals by country of origin[*]

Vít Macháček[1,2]

[1] *vit.machacek@cerge-ei.cz*
CERGE-EI, Politických vězňů 7, 110 00 Prague (Czechia)

[2] Institute of Economic Studies, Faculty of Social Sciences at the Charles University, Opletalova 26, 110 00
Prague (Czechia)

**Abstract**

This study measures the tendency to publish in international scientific journals. For each of nearly 35 thousands Scopus-indexed journals, we derive seven globalization indicators based on the composition of authors by country of origin and other characteristics. These are subsequently scaled up to the level of 174 countries and 27 disciplines between 2005 and 2017. The results indicate that advanced countries maintain high globalization of scientific communication that is not varying across disciplines. Social sciences and health sciences are less globalized than physical and life sciences. Countries of the former Soviet bloc score far lower on the globalization measures, especially in social sciences or health sciences. Russia remains among the least globalized during the whole period, with no upward trend. Contrary, China has profoundly globalized its science system, gradually moving from the lowest globalization figures to the world average. The paper concludes with reflections on measurement issues and policy implications.

**Introduction**

Globalization of science is vital for addressing critical societal challenges (Wagner et al., 2015, 2017). Climate change, access to water or international fishing resources, control of infectious diseases, or social issues linked to development are just a few examples of topics where coordinated action from the international community based on scientific evidence is necessary. Globalization can bring more efficient allocation of research labor across the globe (Gui et al., 2019) and smooth scientific communication with efficient flow of ideas and feedback across borders.

Globalization of science is traditionally approached through *production.* Researchers engaging in international collaboration or academic mobility look for partnerships that help them produce papers more efficiently. We argue for taking a perspective of *scientific communication*, where globalization is derived from researchers' journal submissions' decisions. By sending a paper into a global journal, researchers send a signal that they want to communicate their paper with the whole world. On the contrary, submitting a paper into a journal with only local relevance indicates researchers' plan to communicate with a local audience.

Globalization of scientific communication (from now on just *globalization*) has been shown to be high in Advanced countries in Western Europe and North America (Gazni, 2015; Zitt & Bassecoulard, 1999; Zitt et al., 1998). Researchers in these countries by default publish in international journals. However, much more diverse picture will occur when looking beyond the most developed research systems in the world. In countries of the former Soviet bloc, there is a long history of local publishing which seems to survive to this day (Kirchik et al., 2012; Moed et al., 2018; Pajić, 2015). Local journals have an important role also in China (Zhang et al., 2021), Brazil (Brasil, 2021; Leta, 2012) or in Colombia (Chavarro et al., 2017).

[*] Financial support from the Czech Academy of Sciences for the R&D&I Analytical Centre (RaDIAC) and from the Charles University Grant Agency (GA UK; project no 1062119) is gratefully acknowledged, as well as Elsevier for providing extended access to Scopus API. Earlier versions of the paper were presented at CWTS QSS Internal Meeting and on ISSI 2019 conference in Rome. I thank the participants at these events for their useful comments and suggestions. I also thank Martin Srholec, Ludo Waltman and Nees Jan van Eck as well as an anonymous reviewers for their guidance in the process of writing. All usual caveats apply.

The quantitative literature on the role of international journals in current research system is surprisingly scarce. Up to our knowledge, the only paper analyzing globalization patterns across countries is Zitt & Bassecoulard (1999), published more than 20 years ago. Since then, the global research landscape changed dramatically. It has grown both in size and interconnectedness (Royal Society, 2011; Science Europe, 2013) and collaboration distances increase (Waltman et al., 2011). International collaboration drives the growth of the research output (Adams, 2012, 2013). Developing countries invest heavily to improve their research infrastructure (Adams, 2013; Bornmann et al., 2015; Gazni et al., 2012; Wagner et al., 2015) and international visibility (Zhou & Glänzel, 2010).

This paper aims to bring fresh evidence on globalization of scientific communication. We use data on 34 964 journals indexed in Scopus and calculated globalization scores in 174 countries, 27 narrow and 4 broad disciplines between 2005 and 2017. In the text of this paper, we will only summarize the most important trends across countries, disciplines, and time. Nevertheless, the research community, policymakers, and other relevant stakeholders can benefit from the full detail of the rich dataset. Readers can take advantage of an interactive app [globalizationofscience.com](globalizationofscience.com) specifically designed for comparisons across countries and disciplines.

We build on Zitt & Bassecoulard (1998, 1999), who suggested journals' internationalization indicators based on bibliometric data and the procedure to derive globalization for the whole country and discipline. In this paper, we follow their procedure: 1) we assess the internationalization of all journals in the dataset, 2) in the scaling step, we analyze all published documents in the given country and discipline using the journal internationalization obtained in the first step.

The following section summarizes the literature on globalization of science, the approaches to its bibliometric measurements, and motivations to publish in international journals. Then, we describe our approach to the measurement of journal internationality and globalization of scientific communication. The data section describes a procedure to collect the data and representativeness issues. The results section and the discussion section contain a high-level summary of cross-country and cross-discipline comparisons. The last section concludes.

**Globalization of science**
Communication drives globalization. Scientific ideas, data, results of experiments and analyses, and feedback from peers abroad can be seamlessly distributed across the globe. Academic journals play a central role in this. They serve as critical communication platforms where ideas, results, and feedback meet. International journals facilitate cross-fertilization of ideas. As (Buela-Casal et al., 2006, p. 46) put it: "journals with wider national representation could increase the diversity of ideas and criticisms and be beneficial to the advancement of knowledge."

Scientific endeavor is increasingly globalized (Royal Society, 2011; Science Europe, 2013). The share of internationally co-authored papers on all papers more than doubled in twenty years (Wagner et al., 2015). The average distance between collaborating authors increased five times between 1980 and 2008 (Waltman et al., 2011). In advanced countries, the growth of international collaboration is responsible for the actual growth of the research output (Adams, 2013).

Yet, the dynamics of globalization of science varies across the globe (Adams, 2012, 2013; Frenken et al., 2009; Gazni, 2015; Gazni et al., 2012; Gui et al., 2019; Robinson-Garcia et al., 2019; Royal Socety, 2011; Science Europe, 2013; Wagner & Jonkers, 2017; Zitt & Bassecoulard, 1999). The internationalization of the research system snowballed in the 1980s and 1990s in Western Europe and North America (Zitt et al., 1998). Since then, international collaboration still grows, and the barrier presented by national borders decays (Hoekman et al., 2010). At the same time, developing countries build new scientific capacities (Adams, 2012, 2013; Royal Society, 2011). The prime example of this are BRICS countries (Bornmann et al., 2015; Zhou & Glänzel, 2010; Zhou & Leydesdorff, 2006).

Globalization patterns also differ across disciplines (Gazni, 2015; Gazni et al., 2012; Royal Society, 2011; Science Europe, 2013; Zitt & Bassecoulard, 1999). However, the cross-discipline comparison can turn out to be challenging through the production-based measures as disciplines differ by the tendency to collaborate in general. Gazni et al. (2012) calculated shares of both multi-national and multi-authored publications across disciplines. They found 49 % multi-national publications and 83 % multi-authored publications in space science. These figures are 10 % and 42 % in social sciences. To a significant extent, the disciplinary differences in international collaboration are driven by the general collaboration patterns and team size (Mattsson et al., 2008; Wagner et al., 2017).

When researchers from multiple countries engage in international collaboration, they explicitly agree to produce science together. Each party offers her services – equipment, data, writing skills, labor, knowledge etc. Eventually they start working together and publish a paper. In this sense, collaboration can be perceived as a form of business deal, where researchers can only participate, if they have something to offer. Arguably, globalization measures based on international collaboration indicate the country's competitive advantage in science (Wagner et al., 2001).

Submitting an article into a journal sends a different signal. Journals' characteristics are a crucial evaluation input. The submission points to researchers' motivations and the reward system she faces. And researchers react strategically to the incentives (de Rijcke et al., 2016; Franzoni et al., 2011). When researchers know they are expected to publish internationally, they will find a way to comply. On the contrary if the incentives favor publishing in venues that are disconnected from the rest of the world knowledge flows, there is a good chance that researchers will adjust the publication patterns accordingly.[1]

The production-based approach to globalization considers only a portion of the research output. Senior and elite researchers are more likely to engage in international collaboration and mobility (Czaika & Orazbayev, 2018). The less international the research is, the smaller fraction of the output enters the analysis. Claiming, for instance, only 10 % of researchers in the country engage in international collaboration neglects the globalization efforts of the remaining 90 %. On the contrary, communication-based measures can consider the entire distribution of the research output by analyzing the journal internationalization of each publication. If one is interested in how globally "ordinary researchers" act, globalization of scientific communication can be helpful, especially in the countries where international collaboration is less common.

---

[1] Obviously, also the international collaboration decision is affected by the incentives. Similarly, it would be naïve to separate globalization of scientific communication from scientific capacities entirely. The above distinction serves primarily to suggest main driving forces behind two interrelated concepts. It is beyond the scope of this paper to distinguish between them.

The literature on local and international journals' role in the current research landscape is surprisingly scarce. In general, local journals seems to be in retreat. Zitt & Bassecoulard (1999) documented the growth of international journals at the expense of local journals in all disciplines and all analyzed countries between 1981 – 1997, with the single exception of Russia.

At the same time, formerly local journals gradually internationalize. Gazni (2015) analyzed the development of WoS-indexed journals published by "national" publishers, defined as those publishing journals with addresses from only one country. Authors from other countries than journals' publishers were marked as foreign. The share of papers with foreign authors grew from 36 % in 1990 to 62 % in 2013. The increase in internationalization was present in all disciplines, although significant interdisciplinary differences persisted in 2011-13. Journals also varied across regions – journals from Western Europe, the Pacific region, and Central Africa had over 70 % of papers with foreign authors, while the figure was below 40 % in Latin America (32 %), Middle East (36 %), Asian countries (46 %).

By submitting a publication to a journal, the researcher attempts to address certain community (Chavarro et al., 2014, 2017). Researchers balance the expected costs with the perceived benefits of publication in given venue. By submitting her work to an international journal, the researcher sends a signal that she wants to present to a global audience. On the contrary, by submitting a publication to a journal whose authors are predominantly local, she deliberately chooses to present to another, more localized community.

Non-international journals can be useful in their own right. Chavarro et al. (2017) offer several explanations why people publish in journals not indexed by standard bibliometric databases. Such explanation can easily be applied also to local journal. They identified three communication functions: 1) training of junior researchers, 2) knowledge bridging between mainstream science and local communities, and 3) publishing topics not well covered by mainstream journals. Ma (2019) adds speed and language of publication as an essential factor.

On the other hand, publications in local journals are likely to be less visible than publications in internationalized journals. Kirchik et al. (2012) analyzed the impact of Russian publications published in Russian journals translated to English to Russian publications in non-Russian journals. The number of citations in the latter group was much higher. The publication visibility goes beyond a simple publication language. It is not sufficient to just publish somewhere and expect that "*quality will prevail on its own*". Arguably, if the researcher wants to maximize visibility, she should choose more international journal.

The term international journal is commonly mixed up with publishing in established Western European and North American journals, partly due to the ongoing dominance of these countries on research in general and partly due to the publishing industry concentration in several publishers (Buela-Casal et al., 2006; Larivière et al., 2015). Authors from peripheral countries often cite fear of discrimination as one reason to publish in non-mainstream journals (Kurt, 2018). Instead of publishing in already established journals, the researchers from peripheral countries might be tempted to create their own communities with publication platforms that can develop in an entire ecosystem of journals. This must be carefully considered when constructing internationalization indicators.

The low internationality may be determined by the research topic. In many topics in social sciences and humanities, the object of the study is embedded in the local environment – consider language studies, history or development studies. Examples outside the social sciences include Chinese medicine (Ma, 2019), rice research (Ciarli & Ràfols, 2019), cassava, palm oil, passion fruit (Chavarro et al., 2017) or polar research. The topic is an important factor explaining lower globalization in specific disciplines. Nevertheless, it is unlikely to be driving force of the cross-country differences within the whole research system or broadly defined disciplines, especially in the large countries with a broad portfolio of research activities.

This paper is about the systemic tendency to publish in national journals. It shows the most flagrant cases of countries where the globalization of scientific communication is symptomatically low, regardless of the discipline. Researchers are affected by the incentives provided by the local research systems (Franzoni et al., 2011), which are primarily formed on the national level (Wagner et al., 2015). The resistance of local subjects towards internationalization policies can hinder globalization efforts (Zitt & Bassecoulard, 2004). A vital channel of this resistance can be through national journals, especially in research evaluation systems that rely on "*quantitative evaluation methods that are implemented in a formulaic and rigid manner*" (Rafols et al., 2016, p. 1). Anecdotal evidence of this can often be seen in the countries of the former eastern bloc (Bekavac et al., 1994; Good et al., 2015; Moed et al., 2018). The analysis of globalization of social sciences in Eastern Europe concludes that "it seems that the policies of (some) EE countries are too formal in stimulating futile publication behavior, aimed primarily at quantity, rather than quality" (Pajić, 2015).

Internationality should never be confused with quality. Bad journals can easily be highly international. On the contrary, neither high quality implies high internationality. The low internationality can be determined politically, culturally, and historically. Accusing Russian nuclear physicists of low-quality science would require much better evidence than low globalization of its scientific communication. There is much space for argumentation, exploration, and interpretation. Nevertheless, if there is systemic publishing in journals that are disconnected from the rest of the world's knowledge flows, the relevant stakeholders should be asking: *Is this what we want*?

**Measuring globalization of scientific communication**

Too often, journal internationalization is a loosely, if at all, defined term (Buela-Casal et al., 2006). There is no recognized boundary between "international" and "non-international" journals. There are many options to determine internationality. Indicators can differ by input, as well as by operationalization. The specific definition matters as varying definitions can lead to different rankings by internationalization. Buela-Casal et al. (2006) suggest creating a composite index to solve the ambiguity issue. We prefer keeping the individual indicators as separate inputs for the subsequent analysis to show that measurement artifacts from individual indicators do not drive the main patterns.

Internationalization indicators can be split into four categories according to its input: 1) user-based indicators analyze the community and feedback of users - journals subscribers, readers, or citers; 2) management-based indicators derive internationalization from the structure of the editorial board and publisher characteristics, and 3) author-based indicators consider the country of origin of authors; and 4) content-based indicators use published content as an input, such as language of publication. Due to the availability of data, this work will mostly rely on author-based indicators.

An appealing strategy to assess the journals' internationalization is to compare the number of contributions from the journal's domicile to the total number of papers. However, the major flaw of this approach is that for journals, and especially international, it is unclear what country is the journals' domicile. Arguably, this should be the country where most important editorial decisions are made in. The editorial board composition can be used to derive journals' domicile. However, especially in highly international journals, such a country may not even exist as strategic decisions are made literally globally.

Gazni (2015) uses publishers' location from Web of Science (similar information is also available in Scopus (2018)) as a proxy for the journal's domicile. However, potential misalignment between the publishers' official headquarters and journals' domicile can cause troubles when applying to the whole journal spectrum. To the best of my knowledge, there is no guidance of how the publishers' country is reported to databases. Especially for large multi-national publishers, the publisher country is sometimes puzzling[2].

Moed et al. (2020, 2021) use an indicator called "Index of National Orientation" that proxies the journal's domicile to the country most often contributing to the journal. The authors argue for it due to its simplicity to implement. However, the indicator does not account for the uneven distribution of research output across countries. When applied to all journals indexed in the database, this can lead to 2 types of distortions: 1) The journal with 20 % American or Chinese authors is not equally international to the journal with 20 % Swedish or Czech authors. The first is only slightly nationally biased (US and China currently account for of more than 15 % of current research output), the latter is strongly nationally oriented (Sweden 1 %, the Czech Republic 0.6 %). 2) Large importance of the USA and China for current research leads to misidentifying these countries as a journals' domicile, just because they are large countries.

Instead of vaguely determining journals' domicile, Zitt & Bassecoulard (1998) suggested that journals' internationalization indicators should compare journals' country distribution and the distribution of the entire discipline. Such indicators can naturally account for differences in the country's research sector size. They used distance-based measures comparing the country structure of the journal with the country structure of the aggregate discipline. When the distance between the two is low, meaning that the country structure of the journal closely resembles that of the whole discipline, then the internationality is high. The further apart these distributions are, the more the journal deviates from the aggregate distribution, the less international the journal is. In other words, in an ideally international journal, a likelihood of getting published is independent of the author's country of origin. The less this holds, the less international the journal is.

Another option is to use concentration measures. They are somewhere in between the distance-based measures and measures based on journals' domicile. Sometimes it is convenient not to consider the global dimension of journal internationality. Consider the *European Journal of Public Policy* as an example. It should be highly international, but simultaneously Chinese and American researchers do not need to be equally represented as EU countries. These measures can, but do not have to, account for the size of the research output.

The rest of this section will present seven internationality indicators assessing each journal's globalization for each year. We included four author-based indicators using a country distribution of authors and one indicator using publication language. One indicator employs publishers' domicile, and one indicator measures institutional concentration. Three different data sources enter the analysis – country distribution of authors, institutional distribution of authors, and languages of publications. The indicators are not perfect, but each is imperfect in a different way. When combined, they yield a robust picture of the development of globalization.

---

[2] In Scopus (2018) there is 2,349 active journals published by Elsevier; out of which 51 % is assigned to Netherlands, where Elsevier has its headquarters, 22 % in the United Kingdom and 16 % in the USA.

**Table 1: Globalization Indicators**

| Indicator | $I_{j,d,y,i}$ calculation | Data Description | Indicator type | Source* |
|---|---|---|---|---|
| Euclidian distance | $\sqrt{\sum (x_{j,y} - m_d)^2}$ | Country | Distance | ZB (1998) |
| | Euclidian distance of journal and discipline country distribution | | | |
| Cosine distance | $\dfrac{\sum (x_{j,y} m_d)}{\sqrt{\sum(x_{j,y}^2) \sum(m_d^2)}}$ | Country | Distance | ZB (1998) |
| | Cosine distance of journal and discipline country distribution | | | |
| GiniSimpson Index | $1 - \sum \dfrac{N_{c,j,y}^2}{(\sum N_{c,j,y})^2}$ | Country | Concentration | Aman (2016) |
| | Gini-Simpson diversity of journal country distribution | | | |
| Largest Contributors Surplus** | $\sum_{c=1}^{3} (x_{c,j,y} - m_{c,d})$ | Country | Concentration /Distance | Own |
| | Surplus of three largest contributing countries over its share in discipline | | | |
| Institutional Diversity** | $\sum_{i=1}^{3} (N_{o,j,y}/T_{j,y})$ | Institutional | Concentration | Own |
| | Share of three largest institutions on all documents | | | |
| English Documents | $\dfrac{N_{ENG,j,y}}{T_{j,y}}$ | Language | Index | BC et al. (2006) |
| | Share of English-written documents | | | |
| Local Authors | $\dfrac{N_{LOCAL,j,y}}{T_{j,y}}$ | Country | Index | ZB (1998) |
| | Share of documents from a journal's domicile | | | |

\* ZB is Zitt and Bassecoulard; BC is Buela-Casal
\*\* the underlying data for these indicators are sorted by descending order. The computation algorithm only considers the three most important

*Euclidian* and *Cosine distance* are distance-based indicators. The difference is that cosine distance is independent of the magnitude (the vector length) and only compares angles. The ratio between each pair of countries is more important than the country's percentage share on all articles in the journal. Euclidian distance directly compares percentage shares. *Gini-Simpson index* was suggested by Aman (2016) as a journal internationalization measure "without a bias against periphery." It is a simple concentration measure used in ecology. *Largest Contributors Surplus* combines concentration and distance-based indicators. *Institutional Diversity* captures journal openness rather than internationality per se, and it uses affiliation data. We have also included a simple share of documents written in English (*English Documents*). The *Local Authors* represent journals' domicile-based indicators.

Table 1 provides a detailed overview of the indicators. For each journal $j$ in the dataset, a set of indicators $i$ was calculated for each year $y$. The calculation is derived separately for each discipline $d$. $I_{j,d,y,i}$ denotes the journal internationalization. $N_{c,j,y}$, $N_{c,d,y}$ and $N_{o,j,y}$ are the number of documents with authors affiliated to the country $c$ or organization $o$, in journal $j$ or discipline $d$, in year $y$. $N_{LOCAL,j,y}$ is the number of documents with authors from the same country as the publisher of journal $j$ in the year $y$. $N_{ENG,j,y}$ is the number of English-written documents in the journal $j$ in year $y$. $T_{j,y}$ denotes the total number of documents in the journal $j$ in year $y$. Note that documents by authors from multiple countries are fully attributed to each country, i.e. $T_{j,y} \leq \sum_c N_{c,j,y}$. The vectors $x_{j,y}$ and $m_d$ represent the country distribution of authors of the journal $j$ and the discipline $d$, in which $x_{c,j,y} = \frac{N_{c,j,y}}{T_{j,y}}$ and $m_{c,d} = \frac{\sum_y N_{c,d,y}}{\sum_y \sum_c N_{c,d,y}}$.

Note that for distance-based indicators, the benchmark distribution $m_d$ is calculated from all available periods. We benchmark the journals' distribution against the distribution of authors in the discipline during the whole analyzed period. Changes in journal distribution will therefore affect only $x_{j,y}$ and not $m_d$. The annual changes of the world trend will be fully attributed on the journals' side and not in the discipline aggregate.

*Aggregation*

In the second stage, the journal-level indicators were aggregated to the level of countries and disciplines. The resulting globalization score $G^S_{c,d,y,i}$ is a weighted average of individual journals scaled between 0 and 1, where 0 is the lowest globalization across all years, countries and disciplines within the indicator and 1 is the highest.

The globalization of science in country $c$, discipline $d$ and year $y$ expressed by an indicator $i$ is calculated from the set of journals $J$ assigned to discipline $d$ as an average of individual journals globalization weighted by the share of documents flowing into the journal:

$$G_{c,d,y,i} = \sum_{j=1}^{J} \frac{N_{c,j,y}}{N_{c,d,y}} I_{j,d,y,i}$$

$\frac{N_{c,j,y}}{N_{c,d,y}}$ is the share of documents with authors from country $c$ in journal $j$ on all documents from the country $c$, discipline $d$ in year $y$, $I_{j,d,y,i}$ is the globalization indicator $i$ of journal $j$ in the discipline $d$ and year $y$.

Subsequently, the aggregated globalization index was standardized between 0 and 1 and converted to an ascending scale to simplify the interpretation of the results:

$$G^S_{c,d,y,i} = \frac{G_{c,d,y,i} - G_i^{min}}{G_i^{max} - G_i^{min}} \alpha_i$$

in which $G_i^{min}$ and $G_i^{max}$ is minimum and maximum value of the indicator $i$ across all years, countries and disciplines and $\alpha_i$ equals -1 for the minimizing indicators (low values for high globalization) and 1 otherwise.

To increase results robustness and decrease volatility, the aggregation was only performed when the authors from the country published in at least 30 journals that published at least 30 documents in a respective year. This leads to gaps in results, particularly in the small disciplines and small countries.

**Data**

The analysis is based on the data from the Scopus citation database. Scopus indexes more journals than Web of Science (SCI-Expanded, SSCI, and A&HCI combined; see (Mongeon & Paul-Hus, 2016)). Hence, Scopus is more likely to contain the more local part of the scientific output in the country.

The data for all 34 964 journals indexed in the Scopus Source List (Scopus, 2018) were downloaded using Scopus API in August 2018. For each journal in each year between 2005 – 2017, we downloaded the country and institutional distribution of authors and the distribution of languages. Data were limited to articles, reviews, and conference papers. Scopus (2018) also contains the journals' publisher country as collected by Scopus.

The Scopus Search API was requested with the following query:

$$ISSN(AAAA\text{-}BBBB) \text{ AND } DOCTYPE(AR \text{ OR } RE \text{ OR } CP) \text{ AND } PUBYEAR = YYYY$$

in which AAAA-BBBB is the journal's ISSN and YYYY is the year. Rather than publication-level data, the aggregate distribution is collected. For each journal each year, we collect the number of articles affiliated to each country, language, and institution.

Scopus (2018) uses Scopus Journal Classification (Scopus, 2019) to assign journals to disciplines. We use *Major Subject Classification* of 27 disciplines (referred to as *narrow disciplines*), *Broad Subject Clusters* of 4 disciplines (life sciences, physical sciences, health sciences, and social sciences; referred to as *broad disciplines*), and an aggregate for all disciplines combined - *All*. The most granular level of Scopus classification – *Scopus Subject Areas* – was neglected due to concerns about representativeness and the risk of journals' false identification (Wang & Waltman, 2016). The broad classification will be stressed in the rest of the paper, but the narrow results are also available in both the interactive application and the downloadable data.

Journal-based discipline classification is a rough brush as it is not possible to assign documents directly to disciplines. Many journals (20 % according to broad and 50 % to narrow classification) have more than one discipline in our dataset. The used methodology entirely attributes all journal documents to all assigned disciplines. This may cause distortion, especially due to prominent interdisciplinary journals that index research from various unrelated disciplines.

The data required only minor cleaning. Approximately 5 % of publications with the undefined country were excluded from the analysis. The same number of publications was also subtracted from the total number of publications in the journal. All dependent territories except Hong Kong were dropped.

The resulting database contains information on 22 million documents from the period 2005-2017. The Scopus indexation grows relatively fast (by an average pace of 4 % per year). We track 1.29 million documents published in 2005 up to 2.09 million in 2017. The growth was generally larger in the first half of the period.

*Representativeness of data*

A major drawback of this analysis is the representativeness of the underlying data. We refer to the *globalization of scientific communication*, but it might be more appropriate to refer specifically to the *globalization of scientific communication in journals indexed by Scopus*. Citation databases may index publications, journals, or authors unevenly across countries, disciplines, and time. The following section will analyze the impact of uneven distribution on results.

Bibliometric databases probably represent a more significant portion of the research output in the countries of scientific core than those at the periphery (Chavarro et al., 2017; Ciarli & Ràfols, 2019; Mongeon & Paul-Hus, 2016). With a reasonable assumption that the international journals are more likely to be indexed than non-international, the results can be interpreted as the upper bound of globalization.

The results are sensitive to Scopus journal-indexation decisions. For example, in 2009, Scopus reacted to criticism by increasing its coverage of social sciences and humanities journals by 39 % (Hicks & Wang, 2011). Longitudinal changes must be therefore interpreted with caution. Year-by-year jumps are not necessarily caused by fundamental changes in the researchers' behavior but are often driven by adding (or removing) journals in the database. Also, long-term changes may be driven by indexing a larger portion of existing journals.

The bibliometric databases cover disciplines unevenly as well. Mongeon & Paul-Hus (2016) report significant under-representation of social sciences and humanities in Scopus and WoS. Uneven representativeness is caused by the coverage of journals within the database. Also, for disciplines relying on other publication venues such as books, the results may be distorted. López Piñeiro & Hicks (2015) showed that significant share of research in Spanish sociology is published in journals not indexed in major bibliometric databases. A significant number of researchers, but even whole research topics might be overlooked. There is not much that can be done about this, but it should be kept in mind when interpreting the results. Again, assuming that the not-indexed part of the research is less globalized, we should interpret the results as the upper estimate of globalization.

**Results**

The computation algorithm yielded globalization scores $G^S_{c,d,y,i}$ in three major dimensions - 174 countries ($c$), 31 disciplines (of which 27 *narrow* and 4 *broad* disciplines) ($d$) and the periods between 2005 - 2017 ($y$). The panel is unbalanced as the rule of publishing in at least 30 journals in a given country, discipline, and year is applied. In 2017, for example, the data were available for 171 countries in the total figures across disciplines, 125 – 155 in *broad* disciplines, and less than 100 countries in 21 out of 27 narrow disciplines. Naturally, the larger the research production in the country and the discipline, the more globalization scores are available. The data coverage grows in time, together with the growth of the research output and indexation of journals by Scopus. The globalization scores $G^S_{c,d,y,i}$ (from now on just scores) are scaled between 0 and 1. 0 always represents the lowest globalization across all years, disciplines, and countries computed using the respective indicator, and 1 represents the highest.

The scores computed by different indicators are relatively strongly correlated (see Table 2). 18 out of 21 correlation coefficients exceed 0.5, and half of the coefficients are higher than 0.65. All of the correlations are statistically significant at 1% level. *Euclidian Distance*, *Cosine Distance, Gini-Simpson Index*, *Largest Contributors Surplus,* and *Institutional Diversity* are highly interrelated. Their correlations are at least 0.64, but also 0.8 is no exception. Note that these indicators intrinsically differ in their nature. They include both distance-based and concentration-based indicators, and *Institutional Diversity* even uses an entirely different dataset. The most representative indicator is *Euclidian Distance* with a correlation coefficient higher than 0.75 with all other indicators except *English Documents*. That is why, by default, we refer to it when not stated otherwise.

Only *English Documents* and *Local Authors* stand slightly aside from the rest of the pack. *English Documents* scores suffer from a highly skewed distribution of underlying internationalization of journals. 80% of journals publish documents almost exclusively in English (at least 95 % documents in English), and almost 90 % of all analyzed documents are in English. Only a tiny portion of data thus drives the results. *Local Authors* use vaguely defined publisher country, as already argued aboveNote that "top five most prolific publishers account for more than 50% of all papers published in 2013" (Larivière et al., 2015). These publishers are likely to publish the most international journals where its domicile is the least meaningful.

**Table 2: Correlation matrix of Globalization scores $G^S_{c,d,y,i}$ across indicators i**

| Indicator | Cosine Distance | Gini-Simpson Index | Largest Contr. Surplus | Institutional Diversity | English Documents | Local Authors |
|---|---|---|---|---|---|---|
| **Euclidian Distance** | **.83** | **.87** | **.93** | **.81** | **.61** | **.75** |
| Cosine Distance | | .64 | .75 | .69 | .47 | .41 |
| GiniSimpson Index | | | .72 | .67 | .64 | .78 |
| Largest Contributors Surplus | | | | .79 | .51 | .67 |
| Institutional Diversity | | | | | .43 | .57 |
| English Documents | | | | | | .61 |

*Pearson correlation coefficients of all available data for each indicator; Source: Scopus; own calculation*

*Country and discipline differences*

Figure 1 depicts total globalization scores in 2017 aggregated across disciplines as measured by Euclidian Distance on the map. The darker the color, the lower globalization.[3] The results show a persistent East-West divide. Globalization is high in Western, Northern and Southern Europe, North America, and Australia. On the contrary, the stronghold of low globalization is in the former Soviet Union – Russia, Kazakhstan, or Ukraine. Almost 30 years after the collapse of the Eastern bloc, the continuing isolation of their science systems is still very apparent.

To a lesser extent, globalization seems to be lower in the BRIICS countries. In China, India, Indonesia, Brazil, and South Africa, globalization is below the world average. Also, other large countries in Asia – Turkey, Iran, Iraq, or Pakistan seem to be publishing in local journals more often. Countries with emerging research infrastructure seem to be prone to lower globalization.

Interestingly, the least developed countries often have highly globalized science systems. In the Democratic Republic Congo, Mauritania or Cameroon, globalization is comparable with Great

---

[3] The colors were mapped to globalization using power-law normalization with γ = 0.6, where min = 0.3 and max = 0.9 to highlight the differences in the middle and upper part of the distribution.

Britain, Netherlands, or Sweden. The same applies to globalization in Nicaragua, Costa Rica, Bolivia, or Surinam in Latin America or Myanmar, Afghanistan, and Vietnam in Asia. The production of science in these countries is often highly dependent on international collaboration. Moreover, the underlying data may be covering these science systems poorly.

**Figure 1: Globalization score $G^S_{c,All,2017,Euclidian\ Distance}$, 2017**

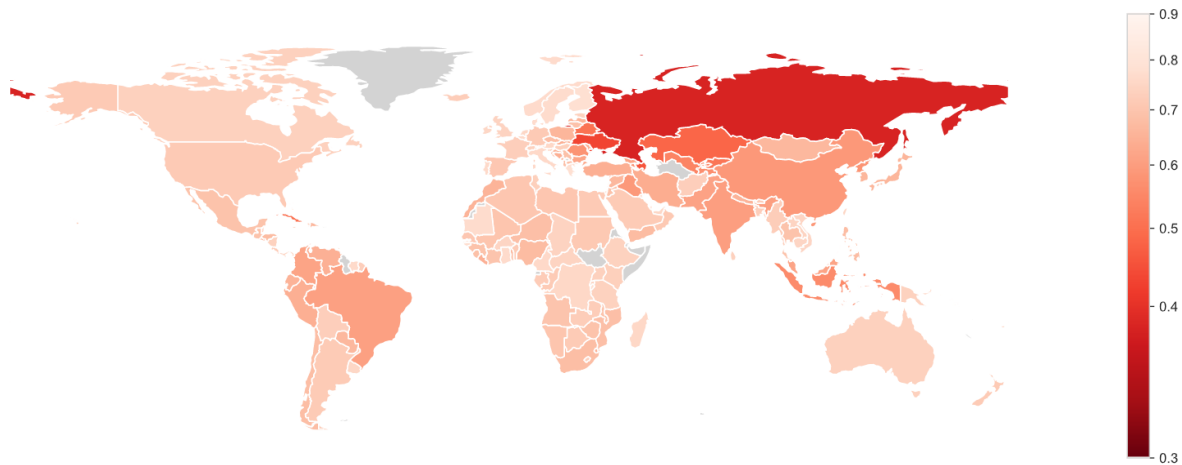

*Note: Globalization scores are assigned to color using Power-Law normalization with $\gamma = 0.6$, where $\min = 0.3$ and $\max = 0.9$. See matplotlib [documentation](#) for details. Grey denotes countries and regions with missing data.*
*Source: own calculation, Scopus.*

The country groups proposed by IMF (2003) are used in more detailed examination by disciplines. The countries are divided into three categories: (1) *Advanced countries* cover the wealthiest countries in the world, mainly in Western Europe, North America, Eastern Asia, Australia, and Oceania. This group should capture the countries of the western core; (2) *Transition countries* consist mainly of the formerly socialistic countries in Central and Eastern Europe and Central Asia, including new EU member states; and (3) *Developing countries* – contain the rest of the world, including China. However, IMF (2003) classification is a rough brush with considerable within-group heterogeneity. Transition countries consist of the Czech Republic or Poland, right next to Kazakhstan, Russia, or Uzbekistan. Similarly, developing countries, for example, contain Saudi Arabia, China, Myanmar, and Bolivia or Democratic Republic Congo. Hence, we further split Transition countries by their EU membership and Developing countries by their continent[4]. The resulting country groups are: 1) *Advanced countries* (32 countries), 2) *Developing – Africa* (49 countries), 3) *Developing – America* (30 countries), 4) *Developing - Asia and Pacific* (35 countries), 5) *Transition – EU* (11 countries) and 6) *Transition - non-EU* (17 countries). The exact mapping of countries to country groups is available in Table A1 in the Appendix.

Figure 2 shows boxen-plots[5] across country groups in 2017. Each section contains distributions of scores from all broad disciplines computed with a respective indicator. Each "boxen" contains scores for different country groups.

---

[4] Only Malta was reassigned from Developing countries to Advanced countries
[5] The line in the middle of each "boxen" represents a median. Each "*box*" of certain width represents a quantile of a certain distance from the median. The more distant quantile from the median, the thinner the box is. Each "boxen" depicts the entire distribution of globalization scores within a group of countries in given years.

The figure confirms the pattern described above. The most globalized are *Advanced countries*. On the opposite side, the least globalized are *non-EU transition countries* – the group consisting mainly of the former USSR and Yugoslavia countries. Second-least globalized are *Transition countries from the EU* – the EU members from the former Eastern bloc followed by *Developing – Asia and Pacific* – fairly diverse group of countries such as *China*, *India*, *Indonesia*, *Turkey* or *Vietnam*.

**Figure 2. Distribution of scores across indicators and country groups, 2017.**

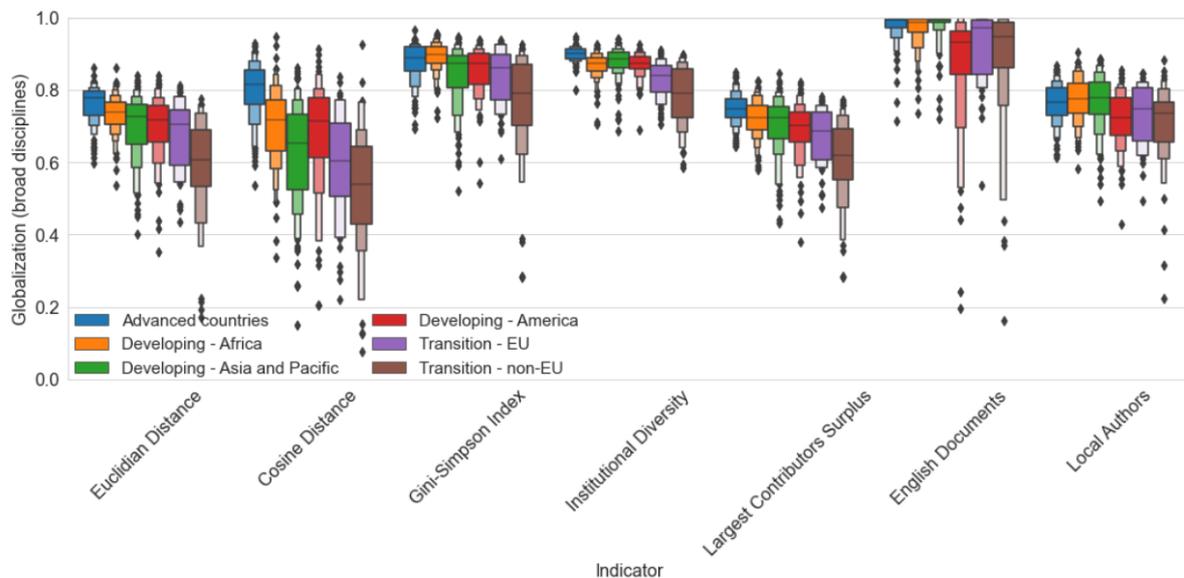

*Note: Each "boxen" contains scores for all broad disciplines within given country group in 2017.*
*Source: own calculation, Scopus.*

Figure 3 shows disciplinary differences across country groups in 2015 - 2017. Each "boxen" represents the distribution of scores in one discipline and one country group. Three years were included to increase the number of observations. Having only one year does not qualitatively change the results. Across country groups, *Life sciences* and *Physical sciences* are more globalized than *Social sciences* and *Health sciences*. What differ across country groups are the gaps, as well as the variance. *Life sciences* and *Physical sciences* are highly globalized in almost all country groups except non-EU transition countries. The differences are minor, if not negligible.

*Social sciences* and *Health sciences* differ across country groups. In *Advanced countries,* they are almost equally globalized to *Life sciences* and *Physical sciences*. The gap between median of *Life* and *Physical* sciences and *Social* and *Health* sciences is wider in other country groups. At the same time, the variance of *Social* and *Health* sciences is larger. For example, in *Social sciences* in *Developing – America* in 2017, Brazil and Cuba are among the least globalized in the world (143[rd] and 140[th] respectively out of 147 countries). The opposite holds for Panama (31[st]) or Jamaica (12[th]).

The large variance and low median apply to all disciplines in *Transition – non-EU* countries. A detailed look reveals that also this group contains two distinct groups – the former USSR countries and the former Yugoslavia. In former USSR countries, scientific communication is strongly isolated from the rest of the world. 8 out of 10 least globalized countries in discipline

All in 2017 belongs to this country group. The Balkan countries' globalization tends to be much higher in all broad disciplines.

**Figure 3: Distribution of scores $G^S_{c,d,y,i}$ across country groups and broad disciplines, 2015-2017**

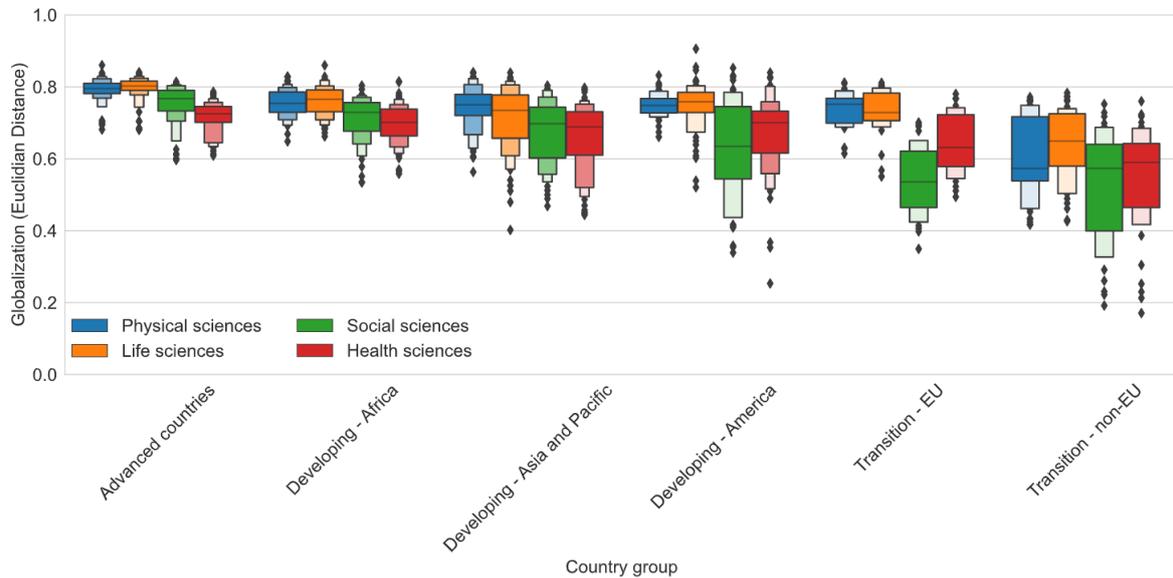

*Note: Each "boxen" contains score for discipline and country group in 2015-2017 computed with Euclidian distance.*
*Source: own calculation, Scopus.*

*Social sciences in the EU*

*Social sciences* stand out in *Transition - EU* countries relative to other disciplines in the same group (see Figure 3). Globalization in this context tends to be much lower than in other disciplines. At the same time, Social Science in the old EU member states (that belong to Advanced countries) is almost equally globalized to *Life* and *Physical sciences*. Hence, we provide a more detailed breakdown of the publication output into journals by internationalization in *Social sciences* in the EU.

**Figure 4: Distribution of documents into journals by globalization quartiles in Social sciences in the EU countries** *(Euclidian distance, 2017)*

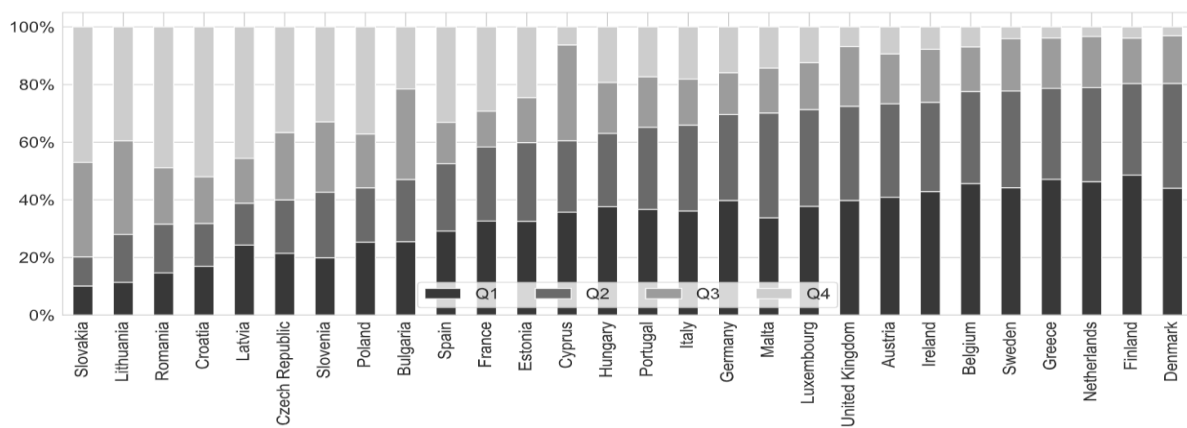

*Note: Q1 represents 25 % of most international journals and Q4 25% least international.*
*Source: own calculation, Scopus.*

First, all journals in *Social sciences* were split into quartiles by their Euclidian distance internationalization in 2017. In the second step, all documents from the given country and discipline were assigned to the journals' quartiles. Quartiles are marked Q1 – Q4, where Q1 is 25% of journals with the highest internationalization and Q4 with the smallest[6]. The darker the color in figure 4, the higher the globalization.

In *Denmark*, *Finland*, *Netherlands*, *Greece*, *Sweden,* or *Belgium,* 80 % of documents were published in journals with above-median internationalization. This figure is only 20 % in *Slovakia*, 28 % in *Lithuania*, 32 % in *Romania*, 39 % in *Latvia,* and 40 % in the *Czech Republic*.

In some Western European countries, researchers rarely publish in the Q4 journals. In 10 countries this figure is below 10 %. In all new member states except Hungary, Estonia, and Cyprus, the Q4 journals accounted for more than 30 % of all documents published in 2017. Among old EU member countries, only France and Spain publish that often in Q4 journals.

*Development over time*
Changes in globalization over time must be interpreted with caution as the results combine changes in the researchers' publication behavior and the indexation decisions on the journals by Scopus.

The globalization scores are surprisingly invariant in time. Figure A1 in the Appendix depicts simple mean scores within given country group and broad disciplines each year between 2005 - 2017. In *Advanced countries* as well as in all *Developing* country groups the globalization does not grow nor decline. During the whole period, the trend in discipline *All, Life sciences* and *Physical sciences* does not indicate anything but stagnation or prolonged growth at best. There are signs of slowly growing globalization in *Transition countries in EU* where globalization steadily grows across disciplines, but it is especially pronounced in Social sciences after 2011.

In detail, we explore 12 large countries. We consider BRIICS countries – *Brazil*, *Russia*, *India*, *Indonesia,* and *South Africa*. These are supplemented by 6 largest countries (other than BRIICS) by a number of documents in 2017 – the *USA*, *United Kingdom*, *Germany*, *Japan*, *France,* and *Italy*, belong to the group of *Advanced countries*. Figure 5 depicts the development of globalization in these countries through the whole analyzed period.

The *Advanced countries* follow the trend of the whole country group. Globalization is high and invariant. The exceptions are the *Social* and *Health sciences* in France and, to a lesser extent, Germany. Countries speaking with what is often dubbed as "world" languages maintain an infrastructure of local journals in *Social* and *Health sciences*. They gradually globalize, but especially in France, many documents are still published locally (see Figure 4). Also, *Japan* seems to be slowly globalizing its research.

---

[6] The closest journal to the median (journal right between Q2 and Q3) as measured by Euclidian distance in Social sciences in 2017 is International Spectator (ISSN 0393-2729), which published 35 documents. 12 had Italian affiliated authors, UK and USA based researchers contributed with 5 documents each. Other countries only had 1 or 2 documents. The closest journal to Q3/Q4 barrier - Zeitschrift fur Slawistik (ISSN 0044-3506) - published 30 documents, out of which 15 had German-affiliated authors, and 5 were Russian. Again, other countries only had 1 or 2 documents.

The case of *Indonesia* (and similarly of *Malaysia*) shows that the path towards higher globalization is not guaranteed. Globalization in these countries declines across disciplines. Further research is needed to ensure that this is not a data artifact of the Scopus' indexation decisions. However, it might be that hand in hand with building national research infrastructure, these countries gradually build their ecosystem of local journals. Over 10 % of Indonesian documents from 2017 were published in the *Journal of Physics: Conference Series*. This journal published almost 30 % of documents with authors from Russia and 12 % of documents had Indonesia-affiliated authors. An additional 8 % of Indonesian research was published in *Advanced Science Letters*, a journal with almost 80 % of authors from Indonesia and Malaysia labeled as "potentially predatory" by J. Beall and whose coverage was discontinued by Scopus after 2017.

**Figure 5: Globalization scores in BRIICS and other large countries over time, Euclidian distance**

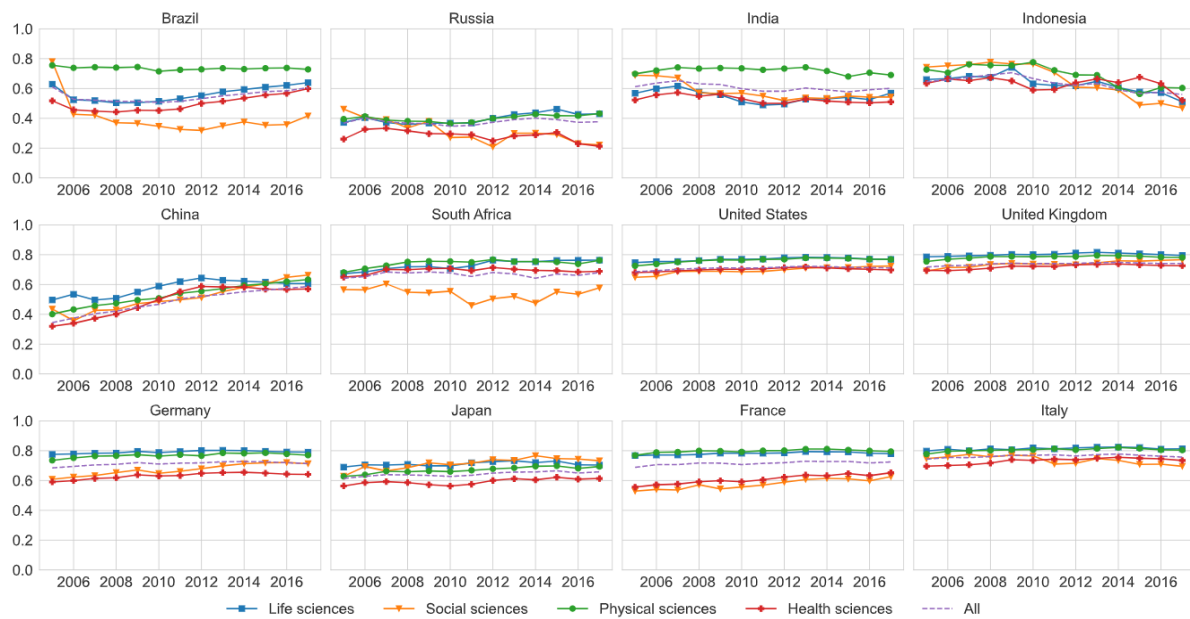

*Source: own calculation, Scopus.*

The strong and persistent isolation of Russia become even more apparent than before. Across all disciplines and indicators, Russia ranks as one of the least globalized countries in the world. Despite government internationalization policies (see Moed et al. (2018)), 90 % of Social Sciences documents, 75 % in Health sciences and almost 65 % in Life and Physical sciences were published in the Q4 journals.

This is in sharp contrast with the case of China. At the beginning of the analyzed period, in 2005, China was the least globalized country in the world according to the total figures and among the five least globalized in all broad disciplines. The rapid transformation of the Chinese system resulted in relatively fast growth of globalization. During the analyzed period, globalization grew fast across all broad disciplines.

The detailed breakdown of the Chinese research output (see Figure 6) reveals that the fast globalization in China can be attributed to the shift of publications from Q4 into the Q2 and Q3 journals. The share of publications in Q1 remains relatively stable in time. At least part of this development can be attributed to the relative growth of Chinese output. While in 2005, Chinese researchers contributed to cca 8 % of all documents in the dataset, the same figure was almost 15 % in 2017. As a consequence, the internationalization scores of Chinese journals can grow not only due to different behavior of Chinese researchers, but also as a result of greater weight that China has as an international actor. That does not mean that Chinese researchers do not increase the presence in the international non-Chinese journals. Nevertheless, its pace is partly driven the total growth of Chinese output.

**Figure 6: Breakdown of research output across disciplines in China between 2005 – 2017** (Euclidian distance)

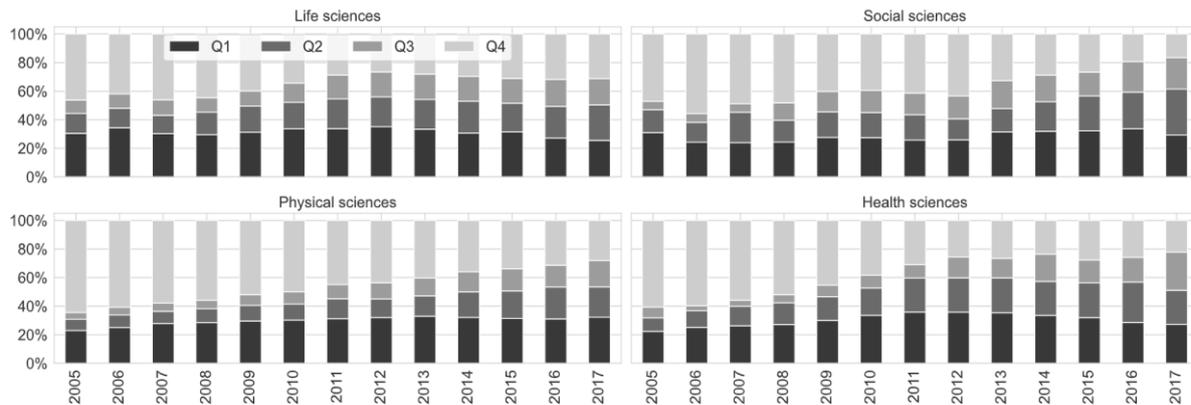

*Source: own calculation, Scopus.*

**Discussion**

Globalization of scientific communication is generally high in the countries of the "core" – in Western Europe and North America. This finding is in line with Zitt et al. (1998), who announced an "*almost complete transition from national to the transnational mode of communication*" already at the end of the last century. Ponds (2009) hypothesized that international collaboration reached its upper limits in western countries like the Netherlands. The space for increasing internationalization is limited if any. Ponds (2009) predicted incorrectly that international collaboration in advanced countries is at its peak. But our data show that the prediction was pretty much correct from the perspective of scientific communication.

The opposite extreme are former USSR countries, where the standard publication model is centered on local journals. There is a long history of publishing in Russian journals (Kirchik et al., 2012). The low globalization is persistent, and it survives policies such as *Project 5-100,* which dedicated special funding to several Russian universities to "jump-off in the ladder" and rank at least five of them in the TOP 100 of respected rankings. Moed et al. (2018) describe an unexpected consequence of Project 5-100 – a fast inflow of Russian national journals into bibliometric databases after the program announcement. It can be challenging to make real change happen.

Doing science in isolated environment is a loss for everyone. Researchers lose visibility, feedback, citations, job opportunities abroad and potential future collaborators. But they are not the only losers. The rest of the world does not have a good access to findings generated within the isolated system. The journals might gain additional prestige if they opened up to the rest of the world. From the systemic perspective, intensive reliance on national journals effectively means that local researchers face lower competition from abroad. Unfortunately, it is likely that researcher lack not only competition, but also an inflow of ideas and feedback.

The systematically low globalization is a form of "free-riding" on the global knowledge flows. Researchers in such systems still have access to mainstream scientific journals' content and can follow what is discussed (at least in the Open-Access journals). However, they do not contribute to it. One option is they have nothing to say, which is only a small step from doing low-quality science. Another option is that they do not want to, for whatever reason. Either way, they fail to promote progress and push knowledge forward regardless of borders. As Louis Pasteur puts it: "*Science knows no country, because knowledge belongs to humanity, and is the torch which illuminates the world.*" (Irving A. Lerch, 1999).

The high globalization of life and physical sciences is intuitive: "*There are no 'American' or 'Russian' electrons, atoms, or galaxies*" (Kirchik et al., 2012). The lower globalization in social sciences, where the object of study is embedded in the local environment, can also be quite natural. Let us take an example of economics. The economy works differently in each country as it is affected not only by the laws of supply and demand but also by institutional factors such as culture and regulation. One side of the discussion (let us label the argument as *contextual*) would stress the necessity to understand local specifics to the inner functioning of the economy. The opposite side – *universalistic* – would argue that the goal of science is to generalize rather than to describe. To isolate the effect of our interest, we need to extract it from its local embedding. We can only do so when we compare different contexts. If two countries differ in globalization within the single discipline, it can be explained in the perspective of *contextual vs. universalistic* debate. Where globalization is low, the researchers favor the *contextual* argument and vice versa researchers in highly globalized countries tend to practice their research in a more *universalist* manner.

Another explanation of the inter-country differences stresses the role of research assessment and the incentives from the local science systems (Franzoni et al., 2011). Country differences are stronger predictors of the resulting globalization than disciplines. The complicated question is what incentives matter and how they translate into higher or lower globalization. Do researchers facing a performance-based research evaluation system react by adjusting their submission decisions (Hicks, 2012)? Are there increased attempts to game the system via indexation of local journals (de Rijcke et al., 2016; Good et al., 2015; Moed et al., 2018)? It is not only about finance (Auranen & Nieminen, 2010; Quan et al., 2017). Successful research transformation requires a sensitive mix of ingredients. Of course, money is important (Franzoni et al., 2011; Quan et al., 2017), but other factors include teaching load, research evaluation requirements, quality of Ph.D programs, mobility or hiring policies (Kuzhabekova & Lee, 2018, 2020; Macháček et al., 2021).

**Conclusions**

The least globalized countries in the world should consider the outcomes of systemic isolation. What is the purpose of having such a strong role of non-international journals? And what are the costs of isolation? If the costs are higher than benefits, how to reform the system efficiently? How to overcome the likely resistance of local subjects and how to support these who are willing to contribute to more open environment? These are all difficult questions that might take years or even decades to resolve successfully.

It cannot be stressed enough that the representativeness of underlying data limits our approach. Assuming that the research beyond the scope of databases is less international than the indexed research implies that the country's globalization is linked to database coverage. The more research is missing from the Scopus database, the more likely it is that real globalization is in fact lower. We can hope that this shortcoming will abate with ongoing professionalization of science, digitalization, and improvement of bibliometric databases. But until then, the caution is warranted. The results are sufficiently robust to summarize major global trends in aggregate and broad disciplines. However, when digging into more detail – the situation in small disciplines and countries – it is essential to remain cautious and combine findings with the contextual information.

An essential motivation behind this paper is informing policies. The rich dataset attached to this paper is ready to be used to deliver policy-oriented studies. Each region, discipline, or country can be analyzed separately and combined with contextual information about the given research system. Hopefully, such studies will deliver comprehensive recommendations and targeted policies.

The literature will benefit from better understanding of how globalization of scientific communication relates to other measures of globalization of science. Various globalization measures – international collaboration, mobility, internationality of citations and of course internationality of journals - should be studied in the cross-country framework. What countries are strongly globalized from the communication perspective, but not that much from the perspective of international collaboration? And vice versa, in which countries researchers often engage in international collaboration, but publishing patterns are unexpectedly locally oriented? Globalization of science is a complex multi-dimensional phenomenon. Detailed understanding of its patterns and identifying the factors driving heterogeneity across indicators will help to better understand the link between research policies and its globalization outcomes.

Better understanding of the complex relationship between globalization and impact, especially in the context of individual career paths would certainly be beneficial. Chavarro et al. (2017) suggests that local journals can serve as "incubators" for junior researchers to learn the publication habits. The globalization dataset can be used to test whether these "learning platforms" are useful or junior researchers would gain (in terms of later impact) from being directly thrown to waters of international publishing.

To address internationalization policies, we need to understand within-country heterogeneity of researchers with respect to globalization. Taking the journals' internationalization scores as an input, it is possible to study intergenerational patterns as well as whether local publishing differ across organizations and disciplines. Two independent clusters of nationally publishing and internationally publishing researchers would require different policies than if researchers rather mix – they publish some articles in international journals and others in local journals.

Another step further is to estimate globalization in higher disciplinary granularity. Using community detection algorithms (Traag et al., 2019), it is possible to identify much more granular disciplines than those of Scopus. This would allow to differentiate between topics that are predominantly linked to international journals and which topics typically end up in more local journals. Such research might help target policies to push internationalization only in topics, in which it is justified, without disrupting research that work best within the local environment.

The fear of discrimination to publish in international journals (Kurt, 2018) must be taken seriously. Globalization of scientific communication means that Eastern researchers publish in Western journals, but ideally, that Western researchers also publish in Eastern journals. In the ideal globalized world, the East-West divide in scientific communication vanishes as researchers from the whole world share their discoveries, knowledge, and feedback in journals that differ by the topic rather than by geographical location. Together they push the frontiers of science forward.

On the contrary, we must seriously ask ourselves whether globalization is always beneficial. Ciarli & Ràfols (2019) documented on the case of rice research that research priority settings do not always meet with the most pressing societal needs in the given the country. The internationalization pressure might be diverging attention of researchers away from the locally important topics that are not considered that important on the international level. Any such indications must be put under careful scrutiny. We must make sure that globalization serve its original purpose - addressing complex societal challenges.

# Appendix

**Table A1: Classification of countries**

| Category | N | List |
|---|---|---|
| Advanced countries | 32 | Australia, Austria, Belgium, Canada, Cyprus, Denmark, Finland, France, Germany, Greece, Hong Kong, Iceland, Ireland, Israel, Italy, Japan, Luxembourg, Netherlands, New Zealand, Norway, Portugal, Singapore, South Korea, Spain, Sweden, Switzerland, Taiwan, United Kingdom, United States, Liechtenstein, Monaco, Malta |
| Developing – Africa | 49 | Algeria, Benin, Botswana, Burkina Faso, Cameroon, Congo, Cote d'Ivoire, Egypt, Ethiopia, Gabon, Ghana, Kenya, Madagascar, Malawi, Mali, Morocco, Mozambique, Namibia, Nigeria, Saudi Arabia, Senegal, South Africa, Sudan, Tanzania, Tunisia, Uganda, Zambia, Zimbabwe, Libya, Gambia, Mauritius, Niger, Togo, Eritrea, Guinea, Rwanda, Swaziland, Lesotho, Angola, Democratic Republic Congo, Sierra Leone, Central African Republic, Seychelles, Mauritania, Guinea-Bissau, Burundi, Liberia, Cape Verde, Chad |
| Developing – Asia and Pacific | 35 | Bangladesh, Cambodia, China, India, Indonesia, Iran, Jordan, Kuwait, Lebanon, Malaysia, Nepal, Oman, Pakistan, Philippines, Sri Lanka, Syria, Thailand, Turkey, United Arab Emirates, Vietnam, Bahrain, Fiji, Iraq, North Korea, Palestine, Qatar, Brunei, Laos, Myanmar, Papua New Guinea, Yemen, Afghanistan, Bhutan, Vanuatu, Solomon Islands |
| Developing - America | 30 | Argentina, Bolivia, Brazil, Chile, Colombia, Costa Rica, Cuba, Ecuador, Guatemala, Jamaica, Mexico, Panama, Peru, Trinidad and Tobago, Uruguay, Venezuela, Barbados, El Salvador, Honduras, Nicaragua, Paraguay, Dominican Republic, Grenada, Haiti, Bahamas, Saint Kitts and Nevis, Dominica, Guyana, Suriname, Belize |
| Transition - EU | 11 | Bulgaria, Croatia, Czechia, Estonia, Hungary, Latvia, Lithuania, Poland, Romania, Slovakia, Slovenia |
| Transition - non-EU | 17 | Armenia, Belarus, Bosnia and Herzegovina, Georgia, Kazakhstan, Macedonia, Mongolia, Russia, Ukraine, Uzbekistan, Azerbaijan, Kyrgyzstan, Moldova, Serbia, Albania, Tajikistan, Montenegro |

**Figure A1: Globalization scores in time across country groups**

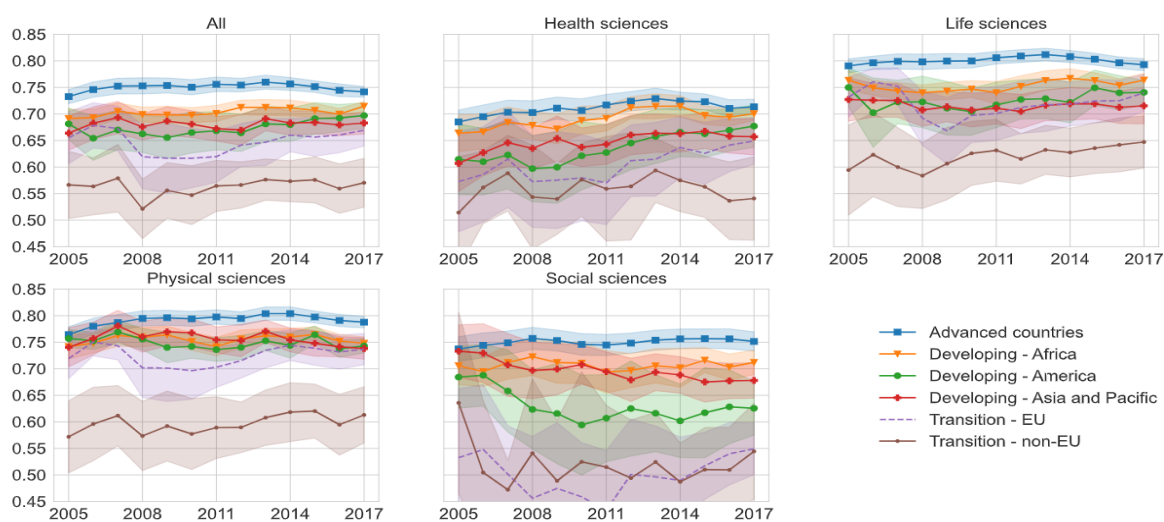

*Note: The plot also includes standard 95 % confidence intervals computed from all scores in given year, country and discipline*

*Source: own calculation, Scopus.*

*Supplementary file Appendix A2: Mean globalization scores for all countries, disciplines, years and indicators*

The CSV file is available at: TBD

*Supplementary file Appendix A3: Distribution of research output into journals by globalization (Euclidian distance) in broad disciplines*

The CSV file is available at: TBD

*Supplementary file Appendix A4: Individual journal globalizations (Euclidian distance, broad set of disciplines, 2017)*

Note that journals with multiple disciplines are computed for each discipline. The results will differ for journals with multiple disciplines for benchmark-based indicators.

The CSV file is available at: TBD